 \documentclass[preprint2]{aastex}

\shorttitle{The orientations of galaxy groups }
\shortauthors{God{\l}owski \& Flin}
 
\begin{document}
 
\title{The orientations of galaxy groups and formation of the Local
Supercluster.}
 
\author{W{\l}odzimierz God{\l}owski}
 
\affil{Uniwersytet Opolski, Institute of Physics, ul.  Oleska  48,
45-052 Opole, Poland}
\email{godlowski@uni.opole.pl}
 
\author{Piotr Flin}
\affil{Jan Kochanowski University, Institute of Physics, ul. Swietokrzyska 15,
25-406 Kielce, Poland}
\email{sfflin@cyf-kr.edu.pl}

\begin{abstract}
We analysed the orientation of galaxy groups in the Local Supercluster (LSC).
It is strongly correlated with the distribution of neighbouring groups in the
scale till about 20 Mpc.  The group major axis is in alignment with both
the line joining the two brightest galaxies  and the direction toward the
centre of the LSC, i.e. Virgo cluster. These correlations suggest that two
brightest galaxies were formed in filaments of matter directed towards the
protosupercluster centre. Afterwards, the hierarchical clustering leads to
aggregation of galaxies around these two galaxies. The groups are formed on
the same or similarly oriented filaments. This picture is in agreement with
the predictions of numerical simulations.
 
\end{abstract}
 
 
\section{Introduction}
 
\citet{b2} was the first who found that major axes of galaxy clusters tend
to point towards their neighbours.  Later the existence of this effect was
discussed by several authors and usually the significant alignment was
reported.
The distance between clusters for which the effect was detected changed from
$10 h^{-1} Mpc$ till $150 h^{-1} Mpc$ (where $h=H_0/100\,km\,s^{-1} Mpc^{-1}$).
The strength of the effect decreases with distance
\citep{s2,f3,r1,u1,p2,w4,c1,h3}. These investigations involved both optical
and X-ray data, as well as clusters belonging (or not) to superclusters.
Nowadays, it is accepted that the effect is not due to selection effects but
is real and its distance scale is between $10 - 60 h^{-1} Mpc$. The alignment
of galaxy group was studied by \citet{w3}.  He used CfA group catalog \citep{g1}
and a catalog based on SSRS ( Southern  Sky Redshit Survey) \citep{m1}. Each
group should have at least $4$ objects and less than $100$. Decontamination
of the groups by foreground and background objects was performed by removing
objects with redshift difference from the group mean over $1000\, km\,s^{-1}$.
They were $59$ groups and Binggeli effect was observed among galaxy groups till
$15- 30 h^{-1} Mpc$. One should note that the investigation of galaxy group
orientation is more difficult than in the case of galaxy clusters. The position
angle of a group consisting of a few objects is determined with much greater
error than for rich clusters. Moreover, statistics with small number are less
reliable. The other investigations of galaxy groups were performed by
\citet{p1}. During study the orientation of $92$ out of $100$ compact groups
listed in the Catalogue of the Compact Groups \citep{h5}, they do not find
alignment, but location of groups along  long chains is  noted.
 
The interpretation of the effect has changed with development of theories,
but the main idea that this should reflect conditions during the structure
formation is still very popular. Numerical simulations gave a better
understanding of physical processes leading to structure formation. These
simulations were performed in the framework of the cold dark matter (CDM) model
presently regarded as the correct description of the large scale structure
formation. Using different approaches and codes, these investigations led to
the conclusion that the preferred orientation of galaxy clusters in the CDM
model is a natural consequence of processes leading to structure formation due
to gravitational interaction. \citet{o1} using large - scale
simulation found  that in  $\Lambda$CDM cosmological model effect reaches
the distance up to $30 Mpc$, while in $\tau$CDM model the range of effects is
twice smaller.  Also in SCDM and OCDM models, where smaller scale simulations
were performed, some alignment effect could be noted.
N - body simulation for standard $\Lambda$CDM model \citep{f1} for $3000$
clusters showed the alignment  of neighbouring clusters  in the distance range
from $10 - 15 Mpc$, while  the Binggeli effect till about $100 Mpc$
is observed. The strong alignment, decreasing with the increasing distance
between clusters, observed till about 100 Mpc was reported \citep{h6}.
The preferred orientation of clusters belonging to superclusters existed also
in SHM + N body simulation . Moreover, from some of the numerical simulations,
it follows that the structure formation occurred along filamentary structures
rather than the walls \citep{f2,s1,b2,h1,h2,a1,w1,w2}.
In order to confirm, or deny, this scheme of structure
origin we carry out an analysis of the Local Supercluster (LSC) galaxy
groups alignment as well as the distribution of the acute angle between the
position angle of the structure and the direction to all remaining clusters.
Groups were taken from Nearby Galaxies (NBG) Catalog \citep{t3}.
We investigated also the alignment of the brightest group galaxy and the parent
group. We determined the position of the line joining the two brightest galaxies
and checked the orientation of this line in respect to parent group and
direction towards Virgo cluster. The distributions of these angles, as well as
differences between some of these angles were tested for isotropy.
 
The paper is organised in the following manner. Section 2 describes
observational data, section 3 presents  statistical method used in the
paper, section 4 presents results and discussions. In section 5 we
formulate conclusions.

\section{Observational data}
 
In the present paper we study the alignment of galaxy groups. Groups were
taken from NBG Catalog \citep{t3}. This Catalog contains 2367 galaxies
with radial velocities less than $3000\,km\,s^{-1}$.
It is complete till magnitude limit $B^{b,i}_T=12.0$  and moreover
dimmer, low-surface brightness gas reach galaxies (late type spirals) are also
incorporated, which ensure that all more massive galaxies are taken into
account. Due to our position in the LSC it is complete for such objects
till the Virgo Cluster centre. An important point
is that this Catalog provides the uniform coverage of entire unobscured
sky \citep{t2}  and only more massive galaxies are taken into account.
Moreover, the the galaxy distances are very  well and in uniform manner
determined. The galaxy distance are based on velocities,
assuming the flat cosmological model ($q_0 =1/2$) with
$H_0 = 75 km \, s^{-1} Mpc^{-1}$ and the model describing velocity
perturbations in the vicinity of the Virgo Cluster \citep{ts1}.
Galaxies position angles were taken from \citet{n1,n2,l1,l2}, and
for 7 missing measurements they were made on PSS prints by the present authors.

The NBG Catalog gives also the affiliation of galaxies to groups.
Groups were found using precise criteria.
In our opinion the groups extracted from the  NBG Catalog \citep{t3}
are  very good observational basis for study of their properties.
 
From the Tully's Catalog we extracted aggregation of galaxies having at
least 40 members, which ensures that at least one substructure has 10
or more members. The substructure having at least 10 members were
taken for the analysis and we call them groups. We decided to select
structures having at least 10 objects, because the determination of structure
great axis is reliable in this case (see also \citet{w4}).
There are $61$ such  groups. Moreover, we repeated the
analysis for subsample containing $35$ groups having at least $20$ galaxies
(Fig.1).
 
It was assumed that groups are two axial ellipsoids. The shape of each group
has been determined considering only the projected position of galaxies on the
celestial sphere in the supergalactic coordinate system $L$, $B$ and applying
the covariance ellipse method. This procedure gives the position angle of the
group major axis.
 
The position angle of each group $PA_g$ is calculated counterclockwise from
the great circle passing through the position of the group centre on the
celestial sphere and the northern pole of the LSC. It was assumed that the
location of a group centre corresponds to the mean of $L$ and $B$ coordinates
of member galaxies and the mean of radial velocity, as given in the Catalog.
Using standard formula from spherical trigonometry, we calculated the
directions between the centre of each group and the centres of the remaining
groups. Each direction is a part of the great circle joining the centres of
two groups. For each group we calculated the acute angle $\phi$ between the
position angle of the major axis of a given group $PA_g$ and direction
towards other groups. In such a manner we have $3660$ directions among groups.
We also investigated the alignment of the brightest group galaxy $PA_{bm}$
and the parent group $PA_g$. Moreover, we determined the position of the
line joining two brightest galaxies in the group $PA_l$ and checked the
orientation of this line relative to the position angle of the parent group
$PA_g$, the position angle of the brightest galaxy $PA_{bm}$ and the direction
towards Virgo cluster $PA_V$. The projected two dimensional line between
two groups is used to compute the acute angle $\phi$

\section{Statistics}
 
\subsection{The general view}
 
We checked for isotropy four discussed distributions of position angles
($PA_g$, $PA_l$, $PA_{bm}$, $PA_V$) having range $0^o$- $180^o$, as well
as differences between position angles
($PA_g-PA_V$, $PA_l-PA_V$, $PA_g-PA_l$, $PA_{bm}-PA_g$, $PA_{bm}-PA_l$,
$PA_{bm}-PA_V$) being the acute angles.
Theoretically, the range of the angles $PA_V$ and $PA_l$ is between
$0^o$ and $360^o$, whereas the range of two remaining angles
$PA_g$ and $PA_{bm}$ is restricted to the range  $0^o$-$180^o$.
We are studding the difference between  direction  of  angles.  In
such analysis, for
$PA_V$ and $PA_l$ an angle $\zeta$ and $ \zeta+180^o$ are identical.
 
The detailed statistical analysis was done using the Fourier test
\citep{h4,f4,k1,g2,g3}, Kolmogorov - Smirnov test
and the $\chi^2$ test. Additionally, we carried out the analysis of
these angles for subsample of 35 groups having at least 20 members.
In all cases the entire range of the analysed angles (position angles and
differences between position angles) was divided into n equal width bins.
In the following analysis we divided the range of
analysed angles into bins of $15$ degrees width, which gives $12$ bins in the
case of position angles and $6$ in the case of differences between position
angles.
Let us denote the total number of analysed groups as $N$, the number of groups
with analysed angles within k-th angular bin as $N_k$ and $N_0$ as mean number
of groups per bin.
 
\subsection{$\chi^2$ test}
 
The dividing range of the analysed angle into $n$ equal width bins,
gives $n-1$ degrees of freedom in the $\chi^2$ test. The value of the
$\chi^2$ statistics is given by the formula
\begin{equation}
\chi^2 = \sum_{k = 1}^n {(N_k -N_{0})^2 \over N_{0}}.
\end{equation}
The $\chi^2$ test yields at the significance level $\alpha=0.05)$ the critical
value of $19.7$ for $11$ degrees of freedom at the significance level
$\alpha=0.05)$ and $11.1$ for $5$ degrees of freedom.

\subsection{The Fourier test}
 
If deviation from isotropy is a slowly varying function of the angle $\theta$
one can use the Fourier test

\begin{equation}
N_k = N_{0,k} (1+\Delta_{11} \cos{2 \theta_k} +\Delta_{21} \sin{2\theta_k})
\end{equation}
where $N_{0,k}$ are expected number of groups per bin (in our case all
$N_{0,k}$ are equal).
 
We obtain the following expression for the $\Delta_{i1}$ coefficients
 
\begin{equation}
\Delta_{11} = {\sum_{k = 1}^n (N_k -N_{0,k})\cos{2 \theta_k} \over
\sum_{k = 1}^n N_{0,k} \cos^2{2 \theta_k}}
\end{equation}
 
\begin{equation}
\Delta_{21} = { \sum_{k = 1}^n (N_k-N_{0,k})\sin{2 \theta_k} \over
\sum_{k = 1}^n N_{0,k} \sin^2{2 \theta_k}}
\end{equation}
with the standard deviation given by expressions:
\begin{equation}
\sigma(\Delta_{11}) = \left( {\sum_{k = 1}^n N_{0,k} \cos^2{2 \theta_k}
} \right)^{-1/2} = \left( {2 \over n N_0} \right)^{1/2},
\end{equation}
\begin{equation}
\sigma(\Delta_{21}) = \left( {\sum_{k = 1}^n N_{0,k} \sin^2{2 \theta_k}
} \right)^{-1/2} = \left( {2 \over n N_0} \right)^{1/2}.
\end{equation}
 
The probability that the amplitude
\begin{equation}
\Delta_1 = \left( \Delta_{11}^2 + \Delta_{21}^2 \right)^{1/2}
\end{equation}
is greater than a certain chosen value is given by the formula
\begin{equation}
P(>\Delta_1 ) = \exp{\left( -{n \over 4} N_0 \Delta_1^2 \right)}
\end{equation}
with standard deviation of this amplitude
\begin{equation}
\sigma(\Delta_1) = \left( {2 \over n N_0} \right)^{1/2}.
\end{equation}
 
This test was originally introduced by \citet{h4} and
substantially modified by \citet{g3} for the case when taking
into account higher Fourier modes:
\footnote{However please note that there is a printed error in
God{\l}owski (1994). Eq. 18 should have form $P(\Delta)=(1+J/2)\exp{(-J/2)}$}
\begin{equation}
N_k = N_{0,k} (1+\Delta_{11} \cos{2 \theta_k} +\Delta_{21} \sin{2
\theta_k}+\Delta_{12} \cos{4 \theta_k}+\Delta_{22} \sin{4\theta_k}+.....).
\end{equation}
 
In our case (all $N_{0,k}$ are equal) it leads to formulas for the
$\Delta_{ij}$ coefficients:
\begin{equation}
\Delta_{1j} = {\sum_{k = 1}^n N_k\cos{2J \theta_k} \over
\sum_{k = 1}^n N_{0} \cos^2{2J \theta_k}},
\end{equation}
and
\begin{equation}
\Delta_{2j} = { \sum_{k = 1}^n N_k)\sin{2J \theta_k} \over
\sum_{k = 1}^n N_{0} \sin^2{2J \theta_k}},
\end{equation}
with the standard deviation
\begin{equation}
 \sigma(\Delta_{11}) = \left( {\sum_{k = 1}^n N_{0} \cos^2{2 \theta_k}
} \right)^{-1/2} = \left( {2 \over n N_0} \right)^{1/2},
\end{equation}
and
\begin{equation}
\sigma(\Delta_{21}) = \left( {\sum_{k = 1}^n N_{0} \sin^2{2 \theta_k}
} \right)^{-1/2} = \left( {2 \over n N_0} \right)^{1/2}.
\end{equation}
 
If we analysed Fourier modes separately, probability that the amplitude
\begin{equation}
\Delta_j = \left( \Delta_{1j}^2 + \Delta_{2j}^2 \right)^{1/2}
\end{equation}
is greater than a certain chosen value is given by the formula:
\begin{equation}
P(>\Delta_j ) = \exp{\left( -{n \over 4} N_0 \Delta_j^2 \right)}.
\end{equation}
 
When we analysed first and second  Fourier modes together  we obtain
\begin{equation}
\Delta = \left( \Delta_{11}^2 + \Delta_{21}^2+\Delta_{12}^2 + \Delta_{22}^2
\right)^{1/2}
\end{equation}
and
\begin{equation}
P(>\Delta ) =\left(1+{n \over 4} N_0 \Delta_j^2 \right) \exp{\left( -{n \over
4} N_0 \Delta_j^2 \right)}.
\end{equation}
 
The value of coefficient $\Delta_{11}$ gives us the direction of departure
from isotropy. If, $\Delta_{11}<0$ then the excess of the group with position
angles near $90$ degrees (parallel to Local Supercluster plane) is observed,
while for $\Delta_{11}>0$ the  excess of the group with position angles
perpendicular to Local Supercluster plane is observed.
 
\subsection{K-S test}
 
The isotropy the resultant distributions of the angles $\phi$ was investigated
using Kolmogorov- Smirnov test (K-S test). We divided our sample according to
distance $D$ between group centres. We assumed that the theoretical, random
distribution contains the same number of objects as the observed one. Our null
hypothesis $H_0$ is that the distribution is random one.  In order to reject
the $H_0$ hypothesis, the value of observed statistics $\lambda$ should be
greater than $\lambda_{cr}$. At the significance level $\alpha=0.05$ the value
$\lambda_{cr} = 1.358$.

\subsection{The linear regression}
 
We study the linear regression  $y=a\phi+b$ between the number of the $\phi$
angles  falling into particular bin and the $\phi$ angle itself for
difference of the $PA_g$ or $PA_l$ and direction towards other groups.

Again we assumed that the theoretical, uniform, random distribution
contains the same number of objects as the observed one. Our null hypothesis
$H_0$  is that the distribution is a random one.  In such a case the
statistics $t=a/\sigma(a)$ has Student's  distribution with $n-2$ degrees
of freedom. We tested $H_0$ hypothesis that $t=0$ against either $H_1$
hypothesis that $t \ne 0$ or, because we expected that decreasing of the
effect with distance, against $H_2$ hypothesis that $t<0$. In order to reject
the $H_0$ hypothesis, the value of observed statistics $t$ should be greater
than $t_{cr}$. We divided the range of analysed angles into $9$ equal bins.
It gives, at the significance level $\alpha=0.05$ the value $t_{cr} =2.365$
in the case of $H_1$ hypothesis and $t_{cr} =1.895$ in the case of $H_2$
hypothesis.

\section{Results and discussions}
 
Firstly, we counted the number of analysed position angles in bins with $20^o$
width and compared them with expected, theoretical distribution.  The quoted
$1\sigma$ error is equal to $\sqrt{N}$, where N is the number of the groups
falling into bin in random distribution. The strong excess of position angles
$PA_g$ is observed in the bin $80^o - 100^o$, which corresponds to the location of
the supergalactic equator. In this bin, the excess of position angles of the
structures is  $5 \sigma$, while the excess of the position line joining two
brightest galaxies ($PA_l$) is $2.5 \sigma$ (for 35 galaxy groups only),
when compared to the number expected in a random distribution.
 
Distributions of the investigated angles for the groups with at least
10 members  are presented in Fig.2 and Fig.3.  Statistics with $PA_{bm}$
were completed using 54 groups for which these angles were taken from
the literature, as well as independently for  sample of 61 groups.
 
The results of the statistical analysis for position angles
($PA_g$, $PA_l$, $PA_{bm}$, $PA_V$) are presented in the Tab.1, where $df$
denotes the number of  degrees of freedom. The distribution of position angles
of the brightest galaxies ($PA_{bm}$) is isotropic, as is observed in galaxy
structures not containing cD galaxy \citep{t0,p0}, which is the case of LSC.
The distribution of group position angles ($PA_g$) is anisotropic
at confidence level $99\%$. The distribution of the line
joining two brightest galaxies ($PA_l$) is anisotropic at the confidence
level $95\%$ only when $35$ richer groups are analysed. The distribution
of the direction towards Virgo centre $PA_V$ is anisotropic at the
confidence level $99\%$.
 
Now we discuss the differences between analysed position angles. Due to
symmetry the range of differences between the position angles
($PA_g-PA_V$, $PA_l-PA_V$, $PA_g-PA_l$, $PA_{bm}-PA_g$, $PA_{bm}-PA_l$,
$PA_{bm}-PA_V$) are restricted to the range between $0^o$ and $90^o$.
The $\chi^2$ test shows that the difference between the group position angle
($PA_g$) and direction towards Virgo cluster ($PA_V$) is not random at the
confidence level $99\%$ (or $95\%$ in the case of poorer groups) (Tab.2).
For $41$ clusters the differences $PA_g-PA_V$ are less than $45^o$, while
only for $20$ clusters they are greater than $45^o$. The distribution of
the difference $PA_V-PA_l$ is anisotropic only for richer groups. The
difference of angles $PA_g-PA_l$ is strongly anisotropic at the confidence
level $99\%$.  The observed excess is below $45^o$. We also performed Fourier
test for differences between position angles. These angles have range only
$90^0$ instead of $180^0$, so only second (and third) Fourier modes can be
taken into account. Therefore, this test can only confirm the existence of
anisotropy.
 
Separately we analysed the differences between $PA_g$ (or $PA_l$) and
direction toward neighbours. These two parameters were chosen because they
describe the possible orientation of structure within the LSC.
The resultant distributions of the angles $\phi$ being the acute angles
between the position angle of the major axis of a given group $PA_g$ and
direction towards other groups for the sample of groups
with at least 10 members are presented in Fig.4.  We divided our sample
according to distance $D$ between group centres.

The results of the statistical analysis are given in Tab.3. At first, we
counted the number of $\phi$ angles in bins with $30^o$ width and compared
them to the expected, theoretical distribution.  The theoretical numbers
falling into bins and errors are rounded to the integer numbers. The first
line of Tab.3 presents the limits  for subsamples division with respect to
the distance $D$ in $Mpc$ between group centres. The next three rows present
the numbers of the $\phi$ -angle in the observed distribution (denoted as
$obs$) falling into three bins and these numbers expected for the theoretical
distribution (denoted as theo) together with their errors. The quoted
$1\sigma$ errors are equal to $\sqrt{N}$,  where $N$ is the number of the
angles falling into bin in random distribution. The greatest deviation
from isotropy (on the $5\sigma$  level) is observed for subsample
containing groups located closer than $10$ Mpc.
The excess of the   $\phi$ - angles is noted in the first bins.
In the case of subsample containing all groups it is $2.9 \sigma$.
Restricting our sample to groups located close each other, with distances
between their centres smaller than $10 Mpc$ the excess is $4.8 \sigma$
diminishing  to $1,4 \sigma$ in the next subsample ($10 <D \leq 20$).
It means that neighbouring groups have tendency to be aligned. This
tendency is vanishing with increasing distance among groups.
The last row of the Tab.3, denoted as K-S, gives the value $\lambda$ of
K-S statistics for each distribution presented in the Fig.4 (and for sample
"all groups"). At the significance level $\alpha = 0.05$
only $2$ of $5$ investigated subsamples are anisotropic. Again, the greatest
anisotropy is observed in the subsample containing groups located closer than
$10$ Mpc.  The distribution of samples $10 < D \leq20$ Mpc
is close to anisotropy. When $D>20$ Mpc distributions are isotropic. The
distribution containing all $61$ groups is anisotropic, which is due to the
subsamples containing closer groups.
The difference between the line joining two brightest galaxies
($PA_l$) and direction towards other groups does not show any clear
evidence for anisotropy.
 
The further  analysis based on linear regression  was performed. We study
the linear regression  $y=a\phi+b$ between the number of the $\phi$ angles
(difference of the $PA_g$ or $PA_l$ respectively, and direction towards
other groups) falling into particular bin and the $\phi$ angle itself.
In Tab.4 we presented the value of $a$ parameter and its error. The analysis
shows that the structures have the tendency to point each other only in the
case when the distance between groups is smaller than $20 Mpc$. This effect
is at the level of $7 \sigma$ for $D \leq 10 Mpc$ and at $2.5 \sigma$ for
$10<D \leq 20 Mpc$. For the sample of $35$ richer groups we observed similar
effect, but at $2.5 \sigma$ level and only for $D \leq 10 Mpc$.
The difference between the line joining two brightest galaxies ($PA_l$)
and direction towards other groups is not uniform at $2.3 \sigma$
level, but only for the sample of "all" structures. This line is in
alignment with the direction toward other groups. The results of linear
regression analysis  are in agreement with that obtained with help of K-S
test but on the higher confidence level. The subsample of richer galaxy
groups confirms these results for both analysed differences.
 
 The above mentioned anisotropies were noted in the coordinate system
connected with LSC. In the equatorial coordinate system, only for difference
between $PA_g$  and directions towards other groups, for subsample
$D \leq 10 Mpc$  the nearly $3 \sigma$  effect was noted.
 
It was pointed out by \citep{s2}
"...if the orientation of galaxy and clusters have a common origin, then the
systems in which galaxy and clusters are aligned ought to be the best
indications of the preferred direction". Moreover, they wrote "...if the
orientations of galaxy and clusters tended to fixed by some pre-existing
direction, then those systems in which galaxy and cluster are aligned outgh
to be the best indicators of the direction, and so we ought to increase the
anisotropy signal relative to the noise by using only the directions of aligned
systems". Therefore we regard the dependence of the group orientation on  the
coordinate system as an additional argument showing the primordial origin of
alignment.

\section{Conclusions}
 
We used two samples of data. The first one contains $61$ groups having at least
$10$ members, in agreement with West's (1989) wish to have such sample of data.
From the second set of data poorer groups were eliminated and $35$ groups
having at least $20$ members remained for analysis. All groups are nearby and
they are located within the Local Supercluster.
 
Our main results are:
\begin{enumerate}
\item The group major axis is in alignment with the line joining the two
brightest galaxies.
\item The group major axis is in alignment with the direction toward the
centre of the LSC.
\item The acute angle between the position angle of the group and direction
 towards each remaining group is not isotropic.
\item The structures have tendency to point each other when the distance
between groups is smaller than $20 Mpc$
 
\end{enumerate}
 
We performed study of the acute angle
between the position angle of the group and direction towards each remaining
group denoted as $\phi$. The sample was divided according to the distance
between group centres. Each subsample was analysed independently using two
coordinate systems. In the equatorial coordinate system the distributions of
the $\phi$ - angle for all subsamples disregarding subsample
$D<10 Mpc$ were isotropic. This is not the case of the coordinate system
connected with the Local Supercluster.
We found that for closer neighbours ($D<10 Mpc$) strong alignment is observed.
It is at almost  $5 \sigma$ level. The Binggeli effect is diminishing
with distance increase, vanishing at about $20 Mpc$. We used the K - S test,
which is usually applied for alignment investigation.
\citet{c2}  criticised it, because it does not point the place,
where the departure of isotropy is observed. They preferred to use the Wilcox
test on rank-sum, which, in their opinion, gave a higher confidence signal for
alignment. However, \citet{o1} applied both tests finding a
little difference in the obtained statistics. Furthermore,  we used the K - S
test only to check  the isotropy of the distribution and not to find the
anisotropy location.
Our results obtained with the help of K-S test are confirmed by linear
regression analysis. The fact that detection of anisotropy is connected with
LSC coordinate system supports the point of view that formation of galaxies
occurred within protostructures. The analysis of the differences between
position angles shows that it is possible that there exists the alignment of
the line joining two brightest galaxies with both position angle of the
parent group and direction towards Virgo cluster centre.
The fact that detection of anisotropy is connected with
LSC coordinate system supports the point of view that formation of galaxies
occurred within protostructures. The analysis of the differences between
position angles shows that it is possible that there exists the alignment of
the line joining two brightest galaxies with both position angle of the
parent group and direction towards Virgo cluster centre. From the presented
analysis of the orientation of galaxy groups in the Local Supercluster the
following picture of the structure formation appears. The two brightest
galaxies were formed first. They originated in the filamentary structure
directed towards the centre of the protocluster. This is the place where
the Virgo cluster centre is located now.
Due to gravitational clustering, the groups are formed in such a manner
that galaxies follow the line determined by the two brightest objects.
Therefore, the alignment of structure position angle and line joining two
brightest galaxies is observed. The other groups are forming on the
same or nearby filament. The flatness of the LSC additionally contributes
to the observed alignment of galaxy groups. The majority of the groups lie
close to us. Due to completeness of the Catalog, the lack of groups
further than the Virgo Cluster centre is observed, but nearby groups are
very well selected and they contain only more massive galaxies.
This picture is in agreement with predictions of several CDM models,
in which structure formation is due to hierarchical clustering. Moreover, the
formation is occurring on the filamentary structure. The further investigation
considering groups clearly  inside  and outside superclusters  on the greater
data set will be very useful to support or reject this picture.

\acknowledgments
 
We thanks the anonymous referee for suggestions and  comments  improving
the paper.
This work was partially supported by the  Jan Kochanowski University
grant BS 052/09.

\clearpage

\begin{figure}
\epsscale{.45}
\plotone{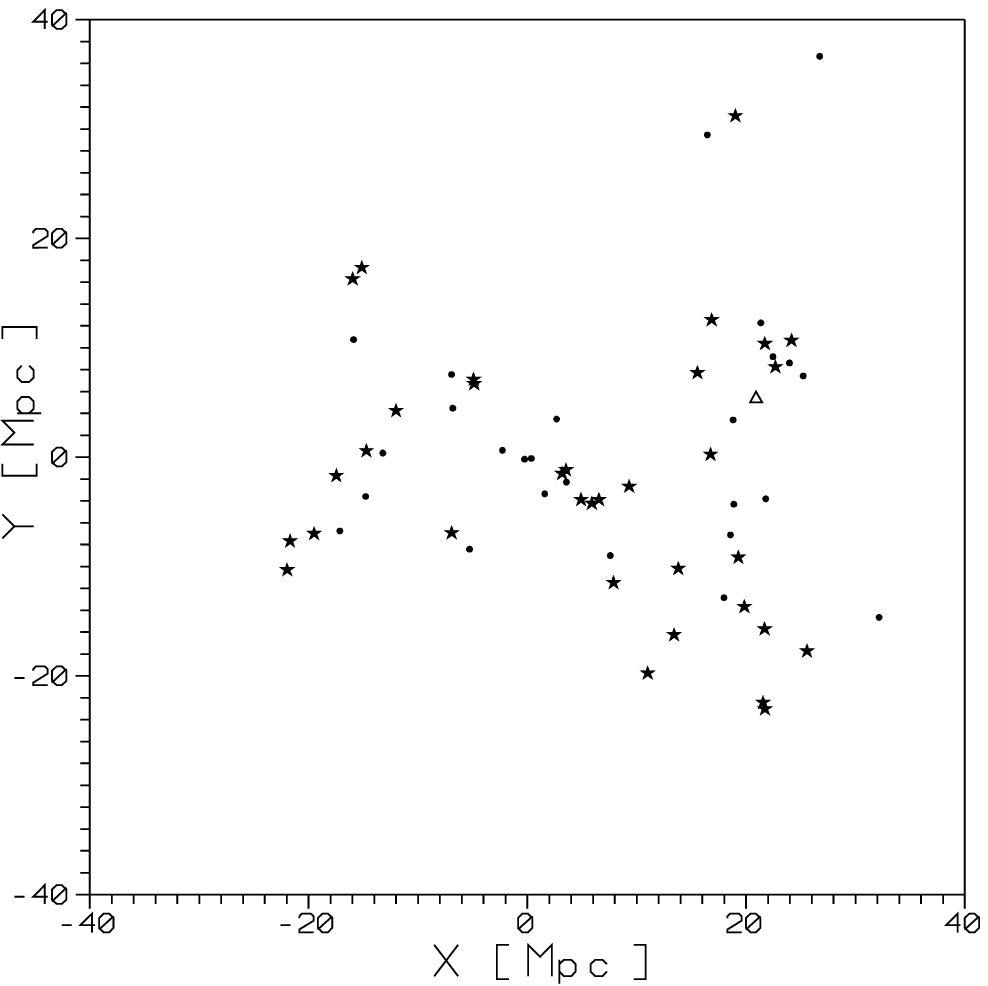}
\plotone{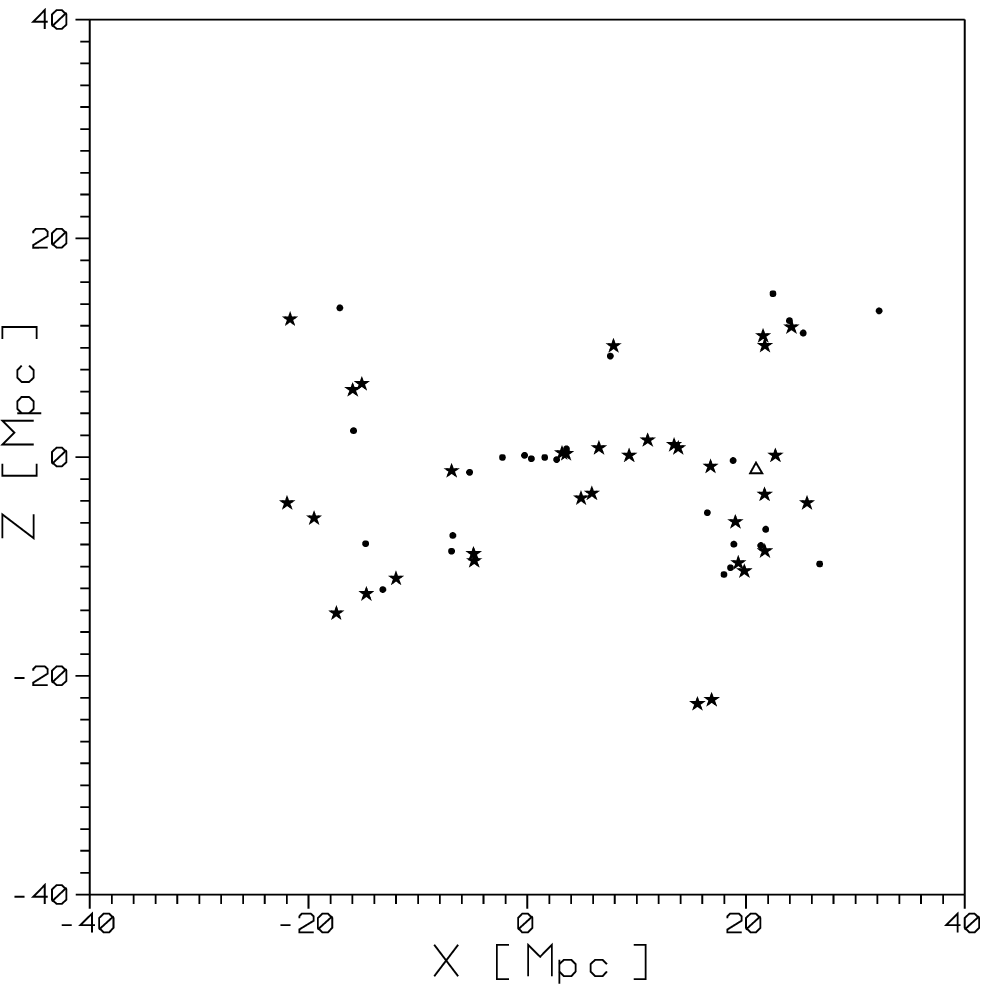}
\caption{The spatial distribution of galaxy groups in the LSC,
Dots - groups with 10- 20 members,  stars - more than 20 member
galaxies,   triangle - Virgo Cluster centre.
On the left panel we present projection into X-Y plane, while on the
right X-Z projection is presented. The coordinate of the Earth are [0,0].
\label{fig1}}
\end{figure}

\begin{figure}
\epsscale{.80}
\plotone{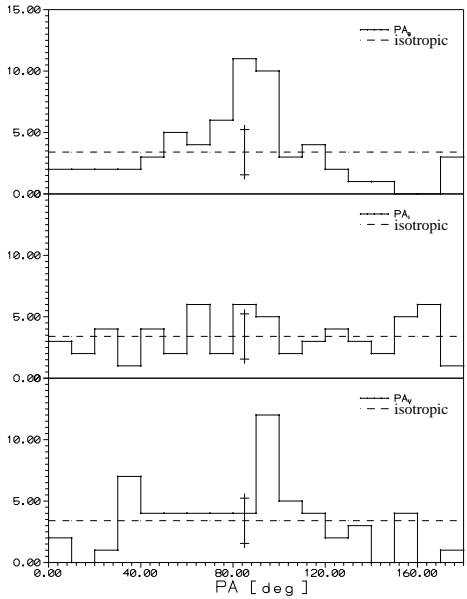}
\caption{The distribution (from top to bottom) of the position
angle of the major axis of a given group $PA_g$, the position of
the line joining two brightest galaxies in the group $PA_l$ and
direction towards Virgo cluster $PA_V$. The dashed line
presents the isotropic distribution. The error bar is equal
$\sqrt{N_0}$ where $N_0$ is average number structures per bin expected in
random distribution.
\label{fig2}}
\end{figure}
 
\begin{figure}
\epsscale{.80}
\plotone{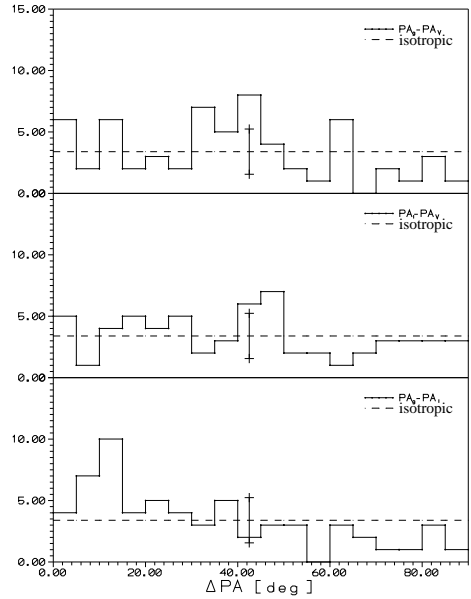}
\caption{The distribution (from top to bottom) of the
differences between position angles
$PA_g-PA_V$, $PA_l-PA_V$, $PA_g-PA_l$.  The dashed line
presentes the isotropic distribution. The error bar is equal
$\sqrt{N_0}$ where $N_0$ is average number structures per bin expected in
random distribution.
\label{fig3}}
\end{figure}
 
\begin{figure}
\epsscale{.72}
\plotone{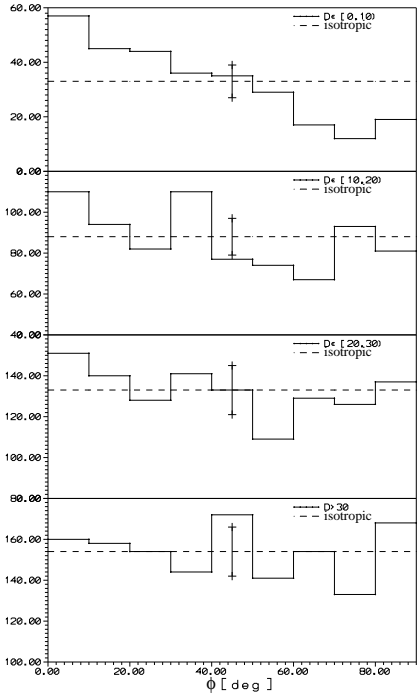}
\caption{The distribution of the angle $\phi$  between the
position angle of the major axis of a given group $PA_g$ and
direction towards other groups. From top to bottom the respectively
distributions for galaxies with $D \leq 10 Mpc$, $10<D\leq 20 Mpc$,
$20<D \leq 30 Mpc$ and $D>30 Mpc$ are presented.  The dashed line
presents the isotropic distribution. The error bar is equal
$\sqrt{N_0}$ where $N_0$ is average number structures per bin expected in
random distribution.
\label{fig4}}
\end{figure}
 
\clearpage
 
\begin{table}
\begin{center}
\caption{Results of the statistical analysis for position angles
\label{Tab.1}}
\begin{tabular}{crrrrrrr}
\tableline\tableline
 
&df&$\chi^2$&$\Delta_{11}/\sigma(\Delta_{11})$&$\Delta_{1}/\sigma(\Delta_1)$&
$\Delta/\sigma(\Delta)$&$P(>\Delta_1)$&$P(>\Delta)$\\
\tableline
$PA_g$    & 11 & 39.1 &-4.89  & 5.20 & 5.85  & $<.001$ & $<.001$\\
$PA_l$    &    &  8.0 &-0.93  & 0.94 & 1.81  &  .644   &  .644\\
$PA_{bm}$ &    &  9.6 &-1.58  & 1.57 & 1.72  &  .290   &  .290\\
$PA_V$    &    & 33.2 &-3.86  & 3.95 & 4.14  & $<.001$ &  .002\\
\tableline
\end{tabular}
\end{center}
\end{table}
 
\clearpage
 
\begin{table}
\begin{center}
\caption{Results of the statistical analysis for differences between position
angles \label{Tab.2}}
\begin{tabular}{crrrrrr}
\tableline\tableline
 
&df&$\chi^2$&$\Delta_{2}/\sigma(\Delta_2)$&$\Delta/\sigma(\Delta)$&$P(>\Delta_2)$&$P(>\Delta)$\\
\tableline
$PA_g-PA_V$    & 5 & 16.0 & 2.20 & 2.71 & .088 & .119\\
$PA_l-PA_V$    &   &  3.4 & 1.61 & 1.62 & .274 & .623\\
$PA_{bm}-PA_V$ &   &  2.7 & 1.17 & 1.47 & .504 & .709\\
$PA_{bm}-PA_g$ &   &  3.8 & 1.84 & 1.93 & .185 & .488\\
$PA_{bm}-PA_l$ &   &  9.9 & 1.84 & 2.34 & .185 & .241\\
$PA_g-PA_l$    &   & 18.4 & 3.46 & 3.96 & .003 & .004\\
\tableline
\end{tabular}
\end{center}
\end{table}
 
\clearpage

\begin{table}
\begin{center}
\caption{Statistical analysis of the distribution of the $\phi$ angle
(between $PA_g$ and direction towards other groups)
\label{Tab.3}}
\begin{tabular}{c|rr|rr|rr|rr|rr}
\tableline\tableline
\multicolumn{1}{c}{}&
\multicolumn{2}{r}{all D}&
\multicolumn{2}{r}{$D\leq 10$}&
\multicolumn{2}{r}{$10<D \leq 20$}&
\multicolumn{2}{r}{$20<D \leq 30$}&
\multicolumn{2}{r}{$D>30$}\\
\tableline\tableline
  & obs & theo & obs & theo & obs & theo & obs & theo & obs & theo \\
\tableline
$0^o - 30^o$  & 1323 &   1220 & 146 &     98 &  286&    263 &  419&    398 &  472&    461 \\
$30^o - 60^o$ & 1201 &$\pm 35$& 100 &$\pm 10$&  261&$\pm 16$&  383&$\pm 20$&  457&$\pm 21$\\
$60^o - 90^o$ & 1136 &        &  48 &        &  241&        &  392&        &  455&        \\
$K-S$ test    &2.276 &        &3.130&        &1.274&        &0.849&        &0.514&        \\
\tableline
\end{tabular}
\end{center}
\end{table}
 
\clearpage
 
\begin{table}
\begin{center}
\caption{The result of the linear regression analysis for $\phi$ angles
\label{Tab.4}}
\begin{tabular}{c|rr|rr|rr|rr|rr}
\tableline\tableline
\multicolumn{1}{c}{}&
\multicolumn{2}{r}{all D}&
\multicolumn{2}{r}{$D \leq 10$}&
\multicolumn{2}{r}{$10<D \leq 20$}&
\multicolumn{2}{r}{$20<D \leq 30$}&
\multicolumn{2}{r}{$D>30$}\\
\tableline\tableline
 
&$a$&$\sigma(a)$&$a$&$\sigma(a)$&$a$&$\sigma(a)$&$a$&$\sigma(a)$&$a$&$\sigma(a)$\\
\tableline
$\phi(PA_g)$&$-1.12$&$\pm 0.26$&$-0.52$&$\pm 0.07$&$-0.31$&$\pm 0.12$&$-0.21$&$\pm 0.14$&$-0.08$&$\pm 0.16$\\
$\phi(PA_l)$&$-0.60$&$\pm 0.26$&$-0.05$&$\pm 0.07$&$-0.21$&$\pm 0.12$&$-0.21$&$\pm 0.14$&$-0.14$&$\pm 0.16$\\
 
\tableline
\end{tabular}
\end{center}
\end{table}
 

\begin{thebibliography}{}
\bibitem[Aragon-Calvo(2007)]{a1} Aragon-Calvo, M. A., van de Weygaert, R.,
Jones, B. J. T., van der Hulst, J. M. 2007, \apj, 655, L5
\bibitem[Basilakos et al.(2006)]{b1} Basilakos, S., Plionis, M., Yepes, G.,
Gottloeber, S., Turchaninov, V. 2006, \mnras, 365, 539
\bibitem[Binggeli(1982)]{b2} Binggeli, B. 1982, \aap, 107, 338
\bibitem[Chambers et al.(2000)]{c1} Chambers, S. C., Melott, A. L., Miller, C. J.
2000, \apj, 544, 104
\bibitem[Chambers et al.(2002)]{c2} Chambers, S. C., Melott, A. L., Miller, C. J.
2002, \apj, 565, 849
\bibitem[Faltenbacher et al.(2002)]{f1} Faltenbacher, A., Gottloeber, S.,
Kerscher, M., Mueller, V.  2002, \aap, 395, 1
\bibitem[Faltenbacher et al.(2005)]{f2} Faltenbacher, A., Allgood, B.,
Gottloeber, S., Yepes, G., Hoffman, Y. 2005, \mnras, 362, 1099
\bibitem[Flin(1987)]{f3} Flin, P. 1987, \mnras,  228, 941
\bibitem[Flin \& God{\l}owski(1986)]{f4} Flin, P.,  God{\l}owski, W. 1986,
\mnras, 222, 525
\bibitem[Geller \& Huhra(1983)]{g1} Geller, M. J., Huhra, J. P. 1983, \apjs,
52, 61
\bibitem[God{\l}owski(1993)]{g2} God{\l}owski, W. 1993, \mnras, 265, 874
\bibitem[God{\l}owski(1994)]{g3} God{\l}owski, W. 1994, \mnras, 271, 19
\bibitem[Hahn et al.(2007a)]{h1}  Hahn, O., Porciani, C., Carollo, C. M.,
Dekel, A. 2007, \mnras, 375, 489
\bibitem[Hahn et al.(2007b)]{h2} Hahn, O., Carollo, C. M., Porciani, C.,
Dekel, A. 2007, \mnras, 381, 41
\bibitem[Hashimoto et al.(2008)]{h3} Hashimoto, Y., Henry, J. P., Boehringer, H.
2008, \mnras, 390,  1562
\bibitem[Hawley \& Peebles(1975)]{h4} Hawley, D. I., Peebles P. J. E. 1975,
\aj, 80, 477
\bibitem[Hickson(1982)]{h5} Hickson, P. 1982, \apj, 255, 382
\bibitem[Hopkins et al.(2005)]{h6} Hopkins, P. F., Bahcall, N. A., Bode, P.
2005 \apj, 618, 1
\bibitem[Kindl(1987)]{k1} Kindl, A. 1987 \aj, 93, 1024
\bibitem[Lauberts(1982)]{l1} Labuerts, A. 1982, ESO/UppsalaSurvey of the ESO B
Atlas, ESO: Garching
\bibitem[Lauberts \& Valentijn(1989)]{l2} Lauberts, A., Valentijn, E. 1989,
The Surface Photometry Catalogue of the ESO-Uppsala Galaxies, ESO:Garching
\bibitem[Maia et al.(1989)]{m1} Maia, M. A. G., da Costa, L. W., Latham, D. V.
1989, \apjs,  69, 809
\bibitem[Nilson(1973)]{n1} Nilson, P. 1973, Uppsala General Catalogue of
Galaxies, Astr. Obs. Ann. V, Vol.1: Uppsala
\bibitem[Nilson(1974)]{n2} Nilson, P. 1974 Catalogue of Selected Non-UGC
Galaxies, Uppsala Astr. Obs. Rep. 5: Uppsala
\bibitem[Onuora \& Thomas(2000)]{o1} Onuora, L.I., Thomas, P.A. 2000, \mnras,
319, 614
\bibitem[Palumbo et al.(1993)]{p1} Palumbo, G. G. C., Saracco, P., Mendes de
Oliveira, C., Hickson, P., Tornatore, V., Baiesi-Pillastrini, G. C. 1993,
\apj, 405, 413
\bibitem[Panko et al.(2009)]{p0} Panko, E., Juszczyk, T., Flin, P. 2009 \aj,
in press
\bibitem[Plionis(1994)]{p2}  Plionis, M. 1994, \apjs, 95, 401
\bibitem[Rhee \& Katgert(1987)]{r1} Rhee, G. F. R. N, Katgert, P. 1987,
\aap, 183, 217
\bibitem[Snedecr \& Cochran(1967)]{sc1} Snedecor, G. W., Cochran W.G. 1967,
Statistical Methods, Iowa University Press: Ames
\bibitem[Springel et al.(2005)]{s1} Springel, V., et al. 2005, \nat, 435, 629
\bibitem[Struble \& Peebles(1985)]{s2} Struble, M. F., Peebles, P. J. E.
1985, \aj, 90, 582
\bibitem[Trevese et al.(1992)]{t0} Trevese, D., Cirimele, G., Flin, P. 1992,
 \aj, 104, 935
\bibitem[Tully \& Shaya(1984)]{ts1} Tully, R. B., Shaya, E. J. 1984, \apj,
281, 31
\bibitem[Tully(1987)]{t2} Tully, R. B. 1987, \apj, 321, 280
\bibitem[Tully(1988)]{t3} Tully, R. B. 1988, Nearby Galaxy Catalog, Cambridge
Univ.Press: Cambridge
\bibitem[Ulmer et al. (1989)]{u1} Ulmer, M., McMillan, S., Kowalski, M. 1989,
\apj, 338, 711
\bibitem[van de Weygaert \& Bond(2008a)]{w1} van de Weygaert, R., Bond, J. R.
2008, A Pan-Chromatic View of Clusters of Galaxies and the Large - Scale
Structures, Plionis, M., Lopez-Cruz, O., Hughes D. Springer: Dordrecht, 335
\bibitem[van de Weygaert \& Bond(2008b)]{w2} van de Weygaert, R., Bond, J. R.
2008, A Pan-Chromatic View of Clusters of Galaxies and the Large - Scale
Structures, Plionis, M., Lopez-Cruz, O., Hughes D. Springer: Dordrecht, 409
\bibitem[West(1995)]{w3} West, M.J., Jones, C., Forman, W.  1995, \apj, 451, L5
\bibitem[West(1989)]{w4} West, M.J. 1989, \apj, 344, 535
\end{thebibliography}
\end{document}